# Search for cascade events with Baikal-GVD

A.D. Avrorin[a], A.V. Avrorin[a], V.M. Aynutdinov[a], R. Bannash[g], I.A. Belolaptikov[b],
V.B. Brudanin[b], N.M. Budnev[c], G.V. Domogatsky[a], A.A. Doroshenko[a],
R. Dvornický[b,h,1], A.N. Dyachok[c], Zh.-A.M. Dzhilkibaev[a], L. Fajt[b,h,i],
S.V. Fialkovsky[e], A.R. Gafarov[c], K.V. Golubkov[a], N.S. Gorshkov[b], T.I. Gress[c],
R. Ivanov[b], K.G. Kebkal[g], O.G. Kebkal[g], E.V. Khramov[b], M.M. Kolbin[b],
K.V. Konischev[b], A.V. Korobchenko[b], A.P. Koshechkin[a], A.V. Kozhin[d],
M.V. Kruglov[b], M.K. Kryukov[a], V.F. Kulepov[e], M.B. Milenin[a], R.A. Mirgazov[c],
V. Nazari[b], A.I. Panfilov[a], D.P. Petukhov[a], E.N. Pliskovsky[b], M.I. Rozanov[f],
E.V. Rjabov[c], V.D. Rushay[b], G.B. Safronov[b], B.A. Shaybonov[*b], M.D. Shelepov[a],
F. Simkovic[b,h,i], A.V. Skurikhin[d], A.G. Solovjev[b], M.N. Sorokovikov[b], I. Stekl[i],
E.O. Sushenok[b], O.V. Suvorova[a], V.A. Tabolenko[c], B.A. Tarashansky[c], and
S.A. Yakovlev[g]

[a]*Institute for Nuclear Research, Moscow, 117312 Russia*
[b]*Joint Institute for Nuclear Research, Dubna, 141980 Russia*
[c]*Irkutsk State University, Irkutsk, 664003 Russia*
[d]*Institute of Nuclear Physics, Moscow State University, Moscow, 119991 Russia*
[e]*Nizhni Novgorod State Technical University, Nizhni Novgorod, 603950 Russia*
[f]*St. Petersburg State Marine Technical University, St. Petersburg, 190008 Russia*
[g]*EvoLogics, Germany*
[h]*Comenius University, Bratislava, Slovakia*
[i]*Czech Technical University in Prague, Prague, Czech Republic*
*E-mail:* dvornicky@dnp.fmph.uniba.sk

Baikal-GVD is a next generation, kilometer-scale neutrino telescope currently under construction in Lake Baikal. GVD is formed by multi-megaton sub-arrays (clusters) and is designed for the detection of astrophysical neutrino fluxes at energies from a few TeV up to 100 PeV. The design of the Baikal-GVD allows one to search for astrophysical neutrinos with flux values measured by IceCube already at early phases of the array construction. We present here preliminary results of the search for high-energy neutrinos via the cascade mode with the Baikal-GVD neutrino telescope.

---

[1]Speaker

1) Introduction

The deep underwater neutrino telescope Baikal-GVD (Gigaton Volume Detector) is currently under construction in Lake Baikal [1]. Baikal-GVD is formed by a three-dimensional lattice of optical modules, which consist of photomultiplier tubes housed in transparent pressure spheres. They are arranged at vertical load-carrying cables to form strings. The telescope has a modular structure and consists of functionally independent clusters - sub-arrays comprising 8 strings. Each cluster is connected to the shore station by an individual electro-optical cable. The first cluster with reduced size, named "Dubna", was deployed and operated in 2015. In April 2016, this array has been upgraded to the baseline configuration of a GVD-cluster, which comprises 288 optical modules attached along 8 strings at depths from 750 m to 1275 m. In 2017-2019, four additional GVD-clusters were commissioned, increasing the total number of optical modules up to 1440 OMs. As Part of phase 1 of the Baikal-GVD construction, an array of nine clusters will be deployed until 2021.

IceCube has discovered a diffuse flux of high-energy astrophysical neutrinos in 2013 [2]. The 7.5 year data sample of high-energy starting events (HESE) comprises 103 events, 77 of which are identified as cascades and 26 as track events [3]. The Baikal Collaboration has long-term experience with the NT200 array to search for diffuse neutrino fluxes via the cascade mode [4,5]. Baikal-GVD has the potential to record already at early phases of construction astrophysical neutrinos with a flux measured by IceCube [6]. One search strategy for high-energy neutrinos with Baikal-GVD is based on the selection of cascade events generated by neutrino interactions in the sensitive volume of the array [7]. Here we discuss the preliminary results based on data accumulated in 2016 and in 2018.

2) Cascade detection with GVD cluster

**2.1 Reconstruction method**

The procedure for reconstructing the parameters of high-energy showers - the shower energy, direction, and vertex - is performed in two steps. In the first step, the shower vertex coordinates are reconstructed using the time information from the telescope's triggered photo-sensors. In this case, the shower is assumed to be a point-like source of light. The $\chi^2$ minimization parameters are shower coordinates ((x, y, z) in a Cartesian coordinate system or $(r,\theta,\varphi)$ in a spherical coordinate system):

$$\chi_t^2 = \frac{1}{(N_{hit}-4)} \frac{\sum_{i=1}^{N_{hit}} \left(T_i(\vec{r}_{sh},t_0)-t_i\right)^2}{\sigma_{ti}^2},$$

where $t_i$ and $T_i$ are the measured and theoretically expected trigger times of the $i$-th photo-sensor, $t_0$ - the shower generation time, $\sigma_{ti}$ - the uncertainty in measuring the time, and $N_{hit}$ is the hit multiplicity. In the case of detecting the Cherenkov radiation of high-energy showers in the Baikal water, the bulk of the photo-sensors are triggered from direct photons or those scattered through small angles. This simplifies considerably the shower coordinate reconstruction procedure. More specifically, the propagation time of the direct photons from the shower to the corresponding photo-sensor can be chosen as the theoretically expected time $T_i$. The reconstruction quality can be increased by applying additional event selection criteria based on the limitation of the admissible values for the specially chosen parameters characterizing the events.

In the second step, the shower energy and direction are reconstructed by applying the maximum-likelihood method and using the shower coordinates reconstructed in the first step. The values of the variables θ, φ, and $E_{sh}$ corresponding to the minimum value of the following functional are chosen as the polar and azimuth angles characterizing the direction and the shower energy:

$$L_A = -\sum_{i=1}^{N_{hit}} \ln p_i(A_i, E_{sh}, \vec{\Omega}_{sh}(\theta, \varphi)).$$

The functions $p_i(A_i, E_{sh}, \Omega_{sh}(\theta,\varphi))$ are the probabilities for a signal with amplitude $A_i$ (measured in photoelectrons) from a shower with energy $E_{sh}$ and direction $\Omega_{sh}$ to be recorded by the *i*-th triggered photo-sensor:

$$p_i = \sum_{n=1}^{\infty} P(n/\bar{n}) \int_{A_i - \frac{\alpha}{2}}^{A_i + \frac{\alpha}{2}} \xi_i(A, n) dA,$$

where $P(n/\bar{n})$ is the probability of detecting *n* photoelectrons at a mean $\bar{n}$ for the Poisson distribution, $\xi(A,n)$ is the probability density function for recording the amplitude *A* at an exposure level of *n* photoelectrons, and α is the scale-division value of the amplitude in photoelectrons. The mean $\bar{n}$ is determined by simulating the responses of the telescope's OMs to the Cherenkov radiation of a shower with energy $E_{sh}$ and direction $\Omega_{sh}$ with allowance made for the light propagation in water, the relative positions and orientation of the OMs and the shower, and the effective OM sensitivity.

## 2.2 Performance of GVD-cluster

The search for high-energy neutrinos with a GVD-cluster is based on the selection of cascade events generated by neutrino interactions in the sensitive volume of the array. Performances of event selection and cascade reconstruction procedures were tested by MC simulation of signal and background events and reconstruction parameters of cascades. After reconstruction of cascade vertex, energy and direction and applying quality cuts, events with a final multiplicity of hit OMs $N_{hit} > 20$ were selected as high-energy neutrino events. The accuracy of cascade energy reconstruction is about 30%, the accuracy of direction reconstruction is about 4 degree (median value) and the vertex resolution is about 2 m [7]. Neutrino effective areas for different flavours averaged over all arrival angles are shown in Fig.1 (left). Energy distributions of cascade events expected for one year observation from astrophysical flux following a power law $E^{-2.46}$ spectrum and single-flavour normalization $4.1 \times 10^{-6}$ GeV$^{-1}$ cm$^{-2}$ s$^{-1}$ sr$^{-1}$ [6, 8], as well as distribution of expected background shower events from atmospheric neutrinos are shown in Fig.1 (right). The expected number of background events from atmospheric neutrinos is strongly suppressed for energies higher than 100 TeV. About 0.6 cascade events per year with energies above 100 TeV and hit multiplicities $N_{hit}>20$ from astrophysical flux following a $E^{-2.46}$ spectrum and 0.08 background events from atmospheric neutrinos are expected.

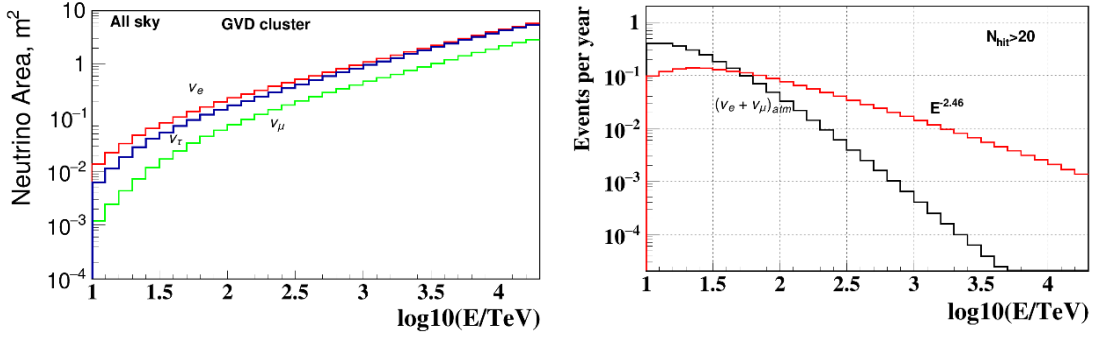

**Figure 1.** Left: Neutrino effective areas for different flavours averaged over all arrival angles. Right: Energy distributions of events expected for one year observation from astrophysical flux with $E^{-2.46}$ spectrum and IceCube normalization (see text). Also shown is a distribution of expected background events from atmospheric neutrinos.

## 3) Data analysis and results

To search for high-energy neutrinos of astrophysical origin the data collected by one cluster in 2016 and by three clusters in 2018 have been used. Life times and sizes of data samples collected by each cluster are shown in Table 1. A data sample of $3.8 \times 10^9$ events has been accumulated by the array trigger, which corresponds to 872 one cluster live days. After applying procedures of cascade vertex and energy reconstruction for hits with charge higher than 1.5 ph.el., 417 cascade-like events with OMs hit multiplicity $N_{hit} > 13$ have been selected. The requirement of high hit multiplicity allows substantial suppression of background events from atmospheric muon bundles. The hit multiplicity dependence on reconstructed cascade energy is shown in Fig.2. 18 events have $N_{hit} > 20$ and 3 of them were reconstructed with energy above 100 TeV and satisfy the requirements for astrophysical neutrino selection. The calculations of the probabilities to obtain such high multiplicity events from atmospheric muons and neutrinos are in progress. The energy and zenith angular distributions of selected events with $N_{hit} > 13$ (green histograms), with $N_{hit} > 20$ (blue histograms) and with $N_{hit} > 20$ and $E > 100$ TeV (red histograms) are shown in Fig.3, left and right panels, respectively. The reconstructed coordinates of selected events are shown in Fig.4. One of three astrophysical neutrino candidates is a contained event. Shown in Table 2 are parameters of the three neutrino candidates: the dates of detection, reconstructed energies, zenith and azimuthal angles as well as distances from cluster vertical axis. The events recorded in 29.04.2016 and in 21.08.2018 are shown in Fig.5 and Fig.6 (left panels). Each sphere represents an OM. Colors represent the arrival times of the photons where red indicates early and blue late times. The size of the spheres is a measure for the recorded number of photo-electrons. In the right panels in Fig.5 and Fig.6 the sky maps of gamma-ray sources as well as a 2° circles around the reconstructed positions of these events are shown.

Table 1: Life time and data sample collected by GVD-clusters in 2016, 2018.

|  | Events | Life time (days) |
|---|---|---|
| Cluster #1, 2018 | 969427196 | 223 |
| Cluster #2, 2018 | 993527274 | 208 |
| Cluster #3, 2018 | 1178496960 | 259 |
| Cluster #1, 2016 | 685523932 | 182 |
| Total | 3826975362 | 872 |

Table 2: Parameters of three cascade events.

|  | Date | Energy, TeV | Zenith, degree | Azimuth, degree | Distance, m |
|---|---|---|---|---|---|
| Cl. #1, 2016 | 29.04.2016 | 155 | 57 | 249 | 45 |
| Cl. #1, 2018 | 21.08.2018 | 153 | 49 | 57 | 77 |
| Cl. #3, 2018 | 24.10.2018 | 107 | 69 | 112 | 89 |

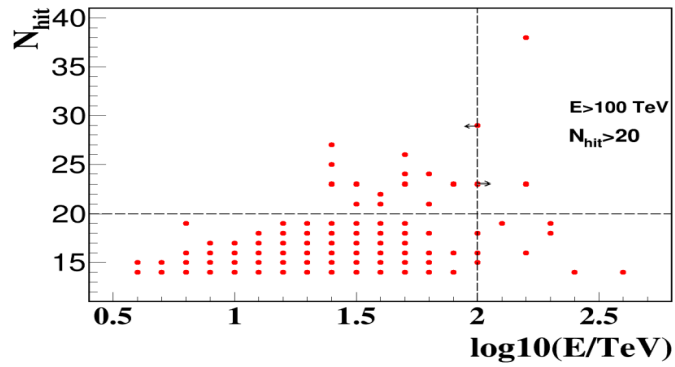

**Figure 2.** Hit multiplicity dependence on the reconstructed cascade energy.

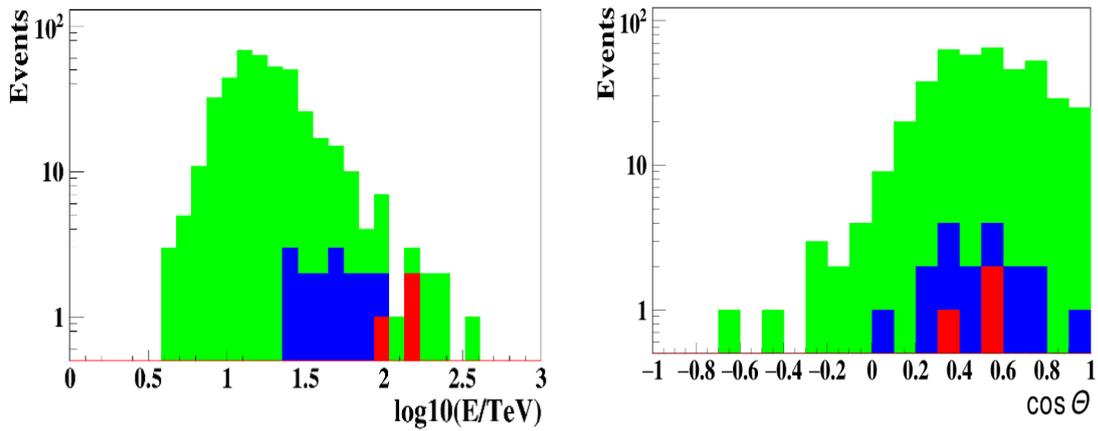

**Figure 3.** Left: Energy distributions of events with $N_{hit} > 13$ (green histograms), with $N_{hit} > 20$ (blue histograms) and with $N_{hit} > 20$ and $E > 100$ TeV (red histograms). Right: Zenith angular distributions of events with $N_{hit} > 13$ (green histograms), with $N_{hit} > 20$ (blue histograms) and with $N_{hit} > 20$ and $E > 100$ TeV (red histograms).

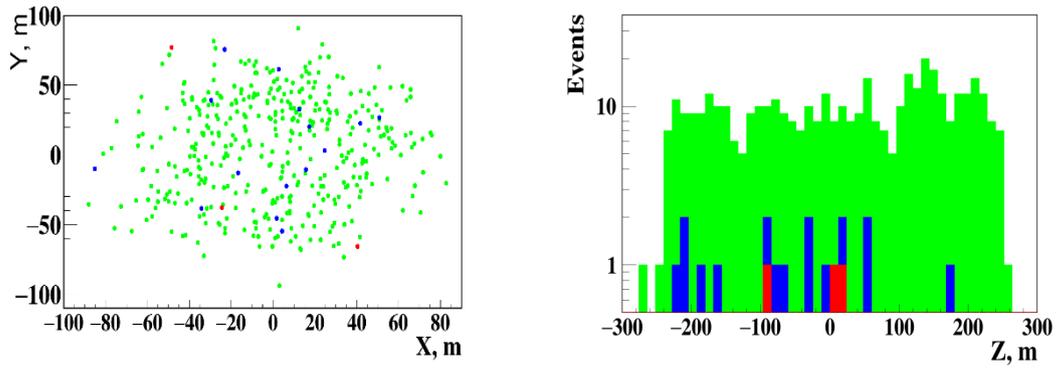

**Figure 4.** Left: Reconstructed cascade vertices in horizontal plane. Green, blue and red dots correspond to events with $N_{hit} > 13$, with $N_{hit} > 20$ and $N_{hit} > 20$ and E>100 TeV. Right: Vertical coordinates of cascade vertices.

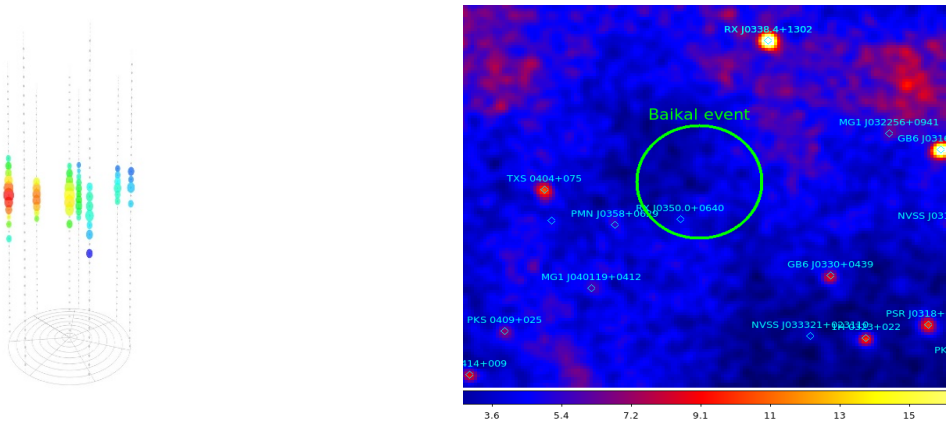

**Figure 5.** Left: The events recorded in 29.04.2016 (see text). Right: The sky map of gamma-ray sources with E > 1 GeV as well as a 2° circle around the reconstructed positions of this event.

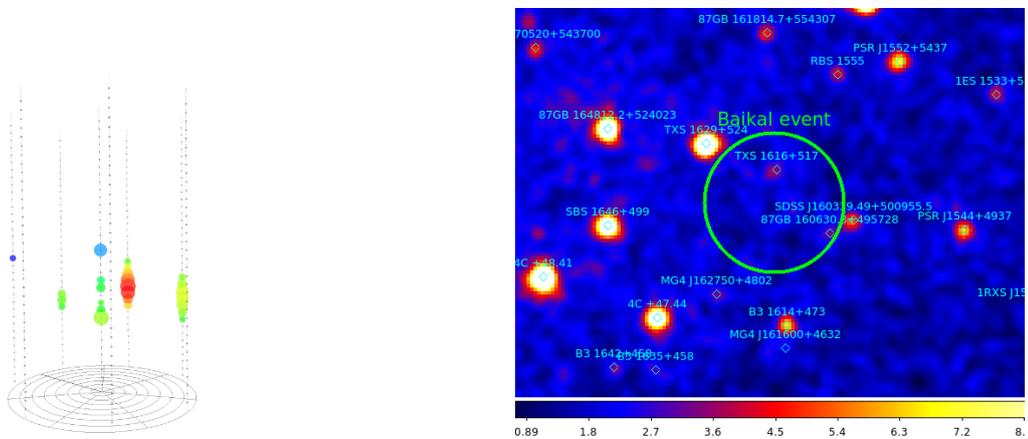

**Fig. 6.** Left: The events recorded in 21.08.2018 (see text). Right: The sky map of gamma-ray sources with E > 1 GeV as well as a 2° circle around the reconstructed positions of this event.

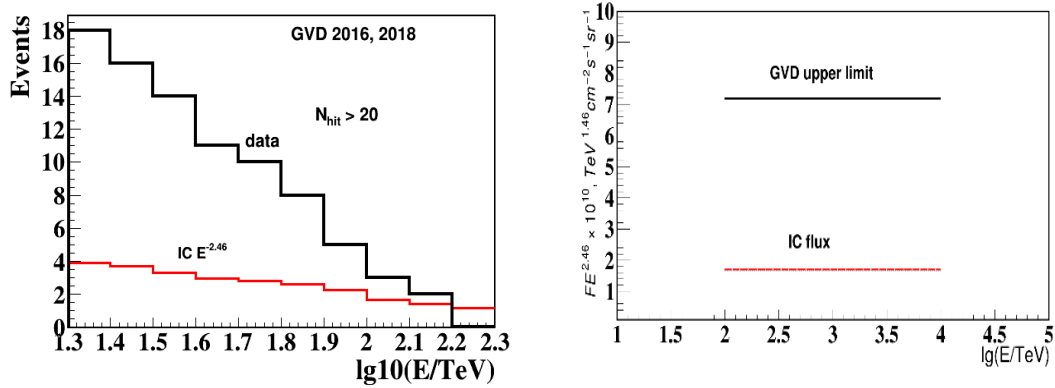

**Figure 7.** Left: Cumulative energy distributions of experimental events (black histogram) and events expected from astrophysical flux with $E^{-2.46}$ spectrum and IceCube normalization (red histogram). Right: 90%CL upper limit on a single-flavor diffuse neutrino flux with $E^{-2.46}$ spectrum and the flux observed by IceCube.

Cumulative energy distributions of experimental events (black histogram) and events expected from IC neutrino flux of astrophysical origin (red histogram) are shown in Fig.7 (left panel). For energies below 100 TeV the data are dominated by background events from atmospheric muons. For energies above 100 TeV higher statistics is required for observation of the astrophysical neutrino flux.

Taking into account the three recorded events with energies above 100 TeV and assuming an $E^{-2.46}$ spectrum single-flavor limit to the flux has been derived according to [9]. This limit is shown in Fig.7 (right), as well as the diffuse flux with IC normalization. Our limit is only a three higher than the IceCube flux, so we are in the same ballpark and, with some more data and a refined analysis, will identify first astrophysical neutrinos.

### 4)   Conclusion

The ultimate goal of the Baikal-GVD project is the construction of a $km^3$-scale neutrino telescope with the implementation of about ten thousand photo-detectors. The array construction was started by the deployment of a reduced-size demonstration cluster named "Dubna" in 2015, which comprised 192 optical modules. The first cluster in its baseline configuration was deployed in 2016. In 2017 and 2018 the second and the third GVD-clusters were deployed. Remarkable achievement has been reached in its 2019 by deployment of two new clusters during one winter expedition.  In total, five clusters with 1140 OMs arranged at 40 strings are data taking since April 2019. Baikal-GVD in 2019 configuration is the largest neutrino telescope in the North at present.  The modular structure of the Baikal-GVD design allows for studies of neutrinos of different origin at early stages of the construction. The analysis of data collected in 2016, 2018 allows the selection of three promising high-energy cascade events - candidates for events from astrophysical neutrinos. The commissioning of the first stage of the Baikal neutrino telescope GVD-1 with an effective volume 0.4 $km^3$ is envisaged for 2020-2021.

*This work was supported by the Russian Foundation for Basic Research (Grants 16-29-13032, 17-02-01237).*